\documentclass[prc,floatfix,nofootinbib,showpacs,showkeys,reprint]{revtex4-1}
\usepackage{graphicx}
\usepackage{amssymb,amsmath,amstext,amsthm,amsfonts}
\usepackage[utf8]{inputenc}
\usepackage{subcaption}
\usepackage{float}
\usepackage{url}

\begin{document}
\today
\title{Muon-neutrino-induced charged-current cross section without pions:\\ Theoretical analysis}

\author{U. Mosel}
\email[Contact e-mail: ]{mosel@physik.uni-giessen.de}
\affiliation{Institut f\"ur Theoretische Physik, Universit\"at Giessen, Giessen, Germany}
\author{K. Gallmeister}
\affiliation{Institut f\"ur Theoretische Physik,\\ Johann Wolfgang Goethe-Universit\"at Frankfurt, Frankfurt a.\ M., Germany}

\begin{abstract}
We calculate the charged current cross sections obtained at the T2K near detector for $\nu_\mu$-induced events without pions in the final state. The method used is  quantum-kinetic transport theory. Results are shown first, as a benchmark, for electron inclusive cross sections on $^{12}$C and $^{16}$O to be followed with a detailed comparison with the data measured by the T2K collaboration on C$_8$H$_8$ and H$_2$O targets. The contribution of 2p2h processes is found to be relevant mostly for backwards angles; their theoretical uncertainties are  within the experimental uncertainties. Particular emphasis is then put on a discussion of events in which pions are first created, but then reabsorbed. Their contribution is found to be essential at forward angles.
\end{abstract}

\maketitle
\section{Introduction}
Pion production, either through resonances or deep inelastic scattering (DIS), is a major process in neutrino-nucleus interactions. At the energies of the Booster Neutrino Beam (BNB) at Fermilab or the T2K experiment it accounts for about 1/3 of the total cross section whereas at higher energy experiments such as MINERvA, NOvA and DUNE it accounts for 2/3 of the total \cite{Mosel:2012kt}. It is thus obvious that any calorimetric method to reconstruct the incoming neutrino energy has to have this channel well under control. Even at the lower energies of the Booster Neutrino Beam (BNB) or T2K, where often a kinematic method is used to reconstruct the energy, pion production cannot be neglected. This is so because this kinematic method relies on an unambiguous identification of quasielastic (QE) scattering. The latter, however, is impossible because the final states of QE scattering (or 2p2h excitations) are always mixed with events in which pions were first created and subsequently reabsorbed inside the target nucleus \cite{Leitner:2010kp}. A complete theory for these reactions thus requires not only a good description of QE and 2p2h processes, but in addition also of pion production and absorption.

About 8 years ago the MiniBooNE  produced the largest data sample for neutrino-induced pion production on nuclei at that time \cite{AguilarArevalo:2010bm}. More recently the experiment MINERvA has also obtained data on pion production, though at a higher energy \cite{Eberly:2014mra,McGivern:2016bwh}. Both of these data sets, the one from MiniBooNE and the one from MINERvA, seem to be incompatible with each other, both in absolute magnitude and in their spectral shape \cite{Sobczyk:2014xza,Alvarez-Ruso:2016ikb}. More recently T2K has also obtained data on pion production, both on CH \cite{Raquel:2015} and on $H_2O$ as targets \cite{Abe:2016aoo} in an energy range close to that of MiniBooNE. We have analyzed the latter data in \cite{Mosel:2017nzk}\footnote{A prediction for the $^{12}$C data can be found in \cite{Mosel:2017zwq}}. There we have shown that the data sets from T2K and from MINERvA, in different energy regimes and on different targets, can be described simultaneously within the same consistent theory. The same is not possible for the MiniBooNE data where a disagreement both with the absolute cross section and the shape of the kinetic energy spectra persists.

Further information on this so-called pion-puzzle may come from events in which pions were first produced and then, subsequently, reabsorbed in the same target nucleus. These processes are contained in so-called 0-pion events in which there are outgoing hadrons, but no pions present in the final state. The cross sections for these 0-pion events thus contain valuable information not only on pion production but also on pion reabsorption and can be used to check the consistency of these two processes. Data on this event-class have been obtained by both MiniBooNE \cite{AguilarArevalo:2010zc} and, more recently, by the T2K ND experiment \cite{Abe:2016tmq,Abe:2017rfw}, in which the incoming energy distribution is similar to, but somewhat narrower, that at the MiniBooNE. Experimentally, the stuck-pion events are indistinguishable from the true QE and 2p2h events. Tuning a generator that does not take these events into account may thus lead to erroneous determinations of other model parameters and then affect extrapolations to new targets and energy ranges.

In the present paper we, therefore, now apply our calculations to the recent T2K data on charged-current 0-pion events on C$_8$H$_8$ \cite{Abe:2016tmq} and H$_2$O \cite{Abe:2017rfw} targets. We will also discuss the theoretical analyses of these data in the experimental papers and a recent theoretical analysis \cite{Megias:2017cuh}.

\section{Method}
For the description of $\nu$-A interactions we use the quantum-kinetic transport-theoretical framework encoded in the GiBUU generator \cite{Buss:2011mx}. This method is based on the non-equilibrium Green's function method \cite{Kad-Baym:1962,Danielewicz:1982kk} and allows to describe a nuclear reaction all the way from the first, initial interaction of the neutrino with the target nucleons to the final state with one outgoing lepton and possibly many outgoing hadrons \cite{Mosel:2016cwa}.

In this theory the reaction is approximately factorized into the initial interaction and the final state interactions (fsi) of the hadrons produced in the initial process. The factorization is only approximate because the initial transition rate and the fsi are linked by the nuclear potential. For example, the outgoing nucleon from an initial QE scattering process experiences a position- and momentum-dependent potential; this affects the initial transition rate. The  nucleon is then propagated onward in exactly the same potential all the way until it leaves the nucleus. In contrast to usually used neutrino generators GiBUU contains a binding potential for all nucleons and allows for off-shell transport. It does not transport the nucleons directly, as in usual Monte Carlo generators, but instead it follows the phase-space distributions of all particles. Pauli-blocking is then handled on the basis of phase-space occupation and not just by sharp momentum cut-offs as in the global Fermi-gas model.

The theory has been described in full detail in Ref.\ \cite{Buss:2011mx} and for its more recent developments in particular for electron- and neutrino-interactions in Ref.\ \cite{Gallmeister:2016dnq}. For easier reference we give in the following subsections some very short descriptions of the treatment of QE scattering, 2p2h excitations and pion production and absorption.

All results shown later in this paper have been obtained with the 2017 version of the GiBUU code which is available for download from Ref.\ \cite{gibuu}. No special tunes or parameter fits have been used; the downloadable version of the code has been used ''out of the box''. The calculations have been made for the target nuclei $^{12}$C and  $^{16}$O since neutrino-induced CC reactions on H cannot contribute to the 0-pion events.

\subsection{QE scattering}
QE scattering depends strongly on the nuclear ground state and the final state potentials that the outgoing nucleon experiences. In GiBUU, starting from a given density distribution, the potential is obtained from an energy-density functional that has been fitted to nuclear matter saturation. Starting from a realistic nuclear density parametrization the potential is calculated. It is not only coordinate dependent, but also explicitly momentum dependent; the momentum-dependence is constrained by fitting proton-nucleus scattering data. For details on these ingredients see Refs.\ \cite{Buss:2011mx} and \cite{Leitner:2006ww}.

The momentum distribution of the groundstate is obtained from the local Fermi-Gas approximation $k_F \propto \rho^{1/3}$. Inserting nucleons with this momentum-distribution into the potential usually leads to a Fermi-momentum that changes over the nuclear volume such that nucleons in the surface region can become unbound.  We have, therefore, in 2016, improved the theory and numerical procedure by requiring the Fermi-energy to be constant over the nuclear volume. This is achieved by iteratively changing density and potential until $k_F$ is constant while still approximately maintaining the local Fermi-Gas connection between density and Fermi-momentum. As shown in \cite{Gallmeister:2016dnq} this leads to a very good description of electron inclusive scattering data and of neutrino data in the QE region.

The calculations do not contain any explicit RPA correlations. In \cite{Martini:2016eec} it was shown that the influence of these correlations is significantly diminished if a bound groundstate, as in the present calculations, is used. This result was recently confirmed in Ref.\ \cite{Nieves:2017lij}.

In all calculations the axial mass parameter is taken to be $M_A = 1$ GeV.

\subsection{2p2h excitations}
Neutrino-nucleus interactions can also take place on two correlated nucleons. This was realized early on by Delorme and Ericson \cite{Delorme:1985ps} and was later on applied to the new set of experiments by Martini et a.\ \cite{Martini:2009uj} and Nieves et al.\ \cite{Nieves:2011pp}. More recently the 2p2h interaction rates have been calculated using a fully relativistic interaction model based on a relativistic free global Fermi-gas \cite{Simo:2016ikv,Barbaro:2017hkz}. In all of these theories the absorption of the neutrino on a pair of nucleons involves excitations of the $\Delta$ resonance. No higher resonance excitations have been taken into account. This limits the applicability of all these theories to the relatively low incoming energies of the BNB experiments and T2K.

In GiBUU, we have, therefore, chosen a different treat\-ment of the 2p2h correlations \cite{Gallmeister:2016dnq} that is free of this limitation and, therefore, also applicable to the higher-energy range of the MINERvA and DUNE experiments. We start with inclusive electron data where a meson exchange contribution (MEC) to the structure function had been extracted from electron-nucleus data by Bosted et al.\ \cite{Bosted:2012qc}. The data set used was characterized by the kinematical constraints: $ 0 < W < 3.2$ GeV, $0.2 < Q^2 < 5$ GeV$^2$; this wide kinematical range can probably never be reached in microscopic calculations. The extraction of this MEC contribution assumed it to be transverse.

To apply this MEC structure function also to neutrino-induced reactions requires two approximations: First, it has to be assumed that also for neutrinos the 2p2h process is predominantly transverse. In light of the recent microscopic calculations \cite{Megias:2016fjk} this is a good approximation. Second, we assume that the reduced vector-vector, axial-axial and vector-axial responses, in which the vector and axial coupling constants and formfactors have been divided out, differ from each other only by kinematical factors (for the actual expressions see \cite{Gallmeister:2016dnq}). We note that this latter approximation underlies also all the work of the Lyon group, starting with the early work of Delorme and M.\ Ericson \cite{Delorme:1985ps} and Marteau \cite{Marteau:1999kt,Marteau:1999jp} and extending up to the more recent work of Martini et al. \cite{Martini:2009uj,Martini:2010ex,Martini:2011wp,Martini:2014dqa}. It finds its theoretical basis in a derivation first given by Walecka et al. in Refs.\ \cite{Walecka:1975,O'Connell:1972zz}.

The success of this treatment of 2p2h interactions has been illustrated in \cite{Gallmeister:2016dnq} where we have shown that  both the MiniBooNE neutrino and antineutrino double differential cross sections can be reproduced without any free parameters or any special tune. The same holds for the T2K inclusive data, both for $\mu$- and $e$-neutrinos.

\subsection{Pion production and absorption}
In the T2K energy regime pions are predominantly produced through the $\Delta$ resonance \cite{Lalakulich:2013iaa}. We use the MAID analysis of electroproduction of pions on nucleons \cite{Drechsel:2007if} as an input for calculations of pion production on Fermi-moving and bound nucleons.  The theory is described in some detail in \cite{Buss:2007ar,Leitner:2008ue}. A validation of the cross sections for photon- and electron-induced reactions is summarized in \cite{Leitner:2009ke} for photon-induced pion production and in \cite{Kaskulov:2008ej} for electroproduction of pions on the nucleus. While MAID fixes the elementary vector couplings and transition form factors, the corresponding axial quantities are obtained from a fit to elementary neutrino-nucleon data \cite{Lalakulich:2010ss}. We use the Argonne-data \cite{Radecky:1981fn} for fixing the free parameters in the resonance and the background amplitudes. The choice of these data is motivated by the reanalysis of the old elementary pion production data that indicated a preference for the Argonne data set \cite{Lalakulich:2010ss,Wilkinson:2014yfa}. As in \cite{Gallmeister:2016dnq} we use free spectral functions without in-medium corrections for the $\Delta$ resonance (see discussion of this point in connection with Fig.\ \ref{fig:C12}).

Pion absorption takes place in GiBUU both by two-nucleon and three-nucleon processes. From experiments with pion beams one knows that both of these processes are essential \cite{Oset:1986yi}. Contrary to other treatments, in GiBUU also the momentum transfer to nucleons is taken into account.

In Ref.\ \cite{Mosel:2017nzk} we have shown that this theory describes the pion production data obtained both at the T2K near detector and at the MINERvA experiment without any free parameters or any special tune.

\section{Testcase: Electron-Nucleus Interactions}
We first compare the predictions of this theory and code with inclusive electron-induced cross sections. This is a necessary check of any such calculation and generator, both of the underlying theory and the numerical implementation. The cross sections for the electron-induced reactions are calculated in GiBUU by using the very same parts of the code, and not some special module, as for the neutrino-induced reactions.

\begin{figure}[h!]
\includegraphics[width=0.7\linewidth,angle=-90] {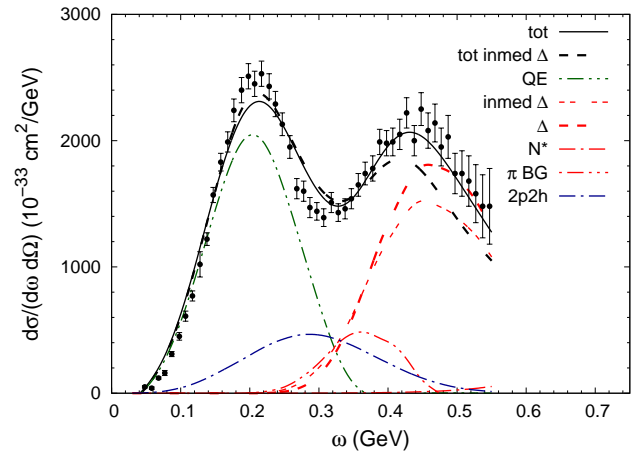}
\caption{Inclusive cross sections for interactions of electrons with $^{12}$C for an incoming energy of 680 MeV and a scattering angle of 60 deg as a function of energy transfer $\omega$. The individual curves give the contributions of the subprocesses QE scattering, 2p2h interactions, $\Delta$ resonance contributions, contributions of higher-lying resonances $N^*$, and the $1\pi$ background con\-tri\-butions.  The curves labeled "inmed $\Delta$" give results of calculations in which the in-medium spectral function of the $\Delta$ resonance from \cite{Oset:1987re} was used. The data are taken from the arXiv for quasielastic electron-nucleus scattering data \cite{Benhar:2006er}.}
\label{fig:C12}
\end{figure}

Since we have already shown a number of comparisons with electron-scattering data for the C target in Ref.\ \cite{Gallmeister:2016dnq}, we show here in Fig.\ \ref{fig:C12} only one example at an energy of 680 MeV, corresponding roughly to the peak energy of the T2K flux.
Figure \ref{fig:C12} shows an excellent agreement with data over the full energy-transfer range.

Two aspects of this result are worthwhile to comment on in some more detail. First, it is seen that the 2p2h contribution, while peaking in the dip region between the QE and $\Delta$ peaks, is already quite sizable even under the QE peak.
Second, the figure contains results obtained with both the free $\Delta$ spectral function and a collision-broadened one from \cite{Oset:1987re}. The results using the free spectral function agree very well with the experiment while the in-medium spectral function of Ref.\ \cite{Oset:1987re} leads to a  cross section that is significantly too low in the $\Delta$ peak region. That the data obviously need the higher cross section at the resonance peak obtained with a free spectral function was already observed in Ref.\ \cite{Gallmeister:2016dnq}. This is consistent with the fact that in the present theory the $\Delta$ resonance is excited only by one-body processes. The width of the resonance contribution, on the other hand, is hardly affected. This is in line with the results of \cite{Lehr:2001ju} where it was shown that an in-medium collisional broadening has only a minor influence on the observable peak width. A much larger effect comes from an additional broadening due to Fermi motion.

\begin{figure*}[ht!] 
	\centering
\begin{subfigure}{0.5\textwidth}
	\centering
\includegraphics[width=0.5\linewidth,angle=-90]{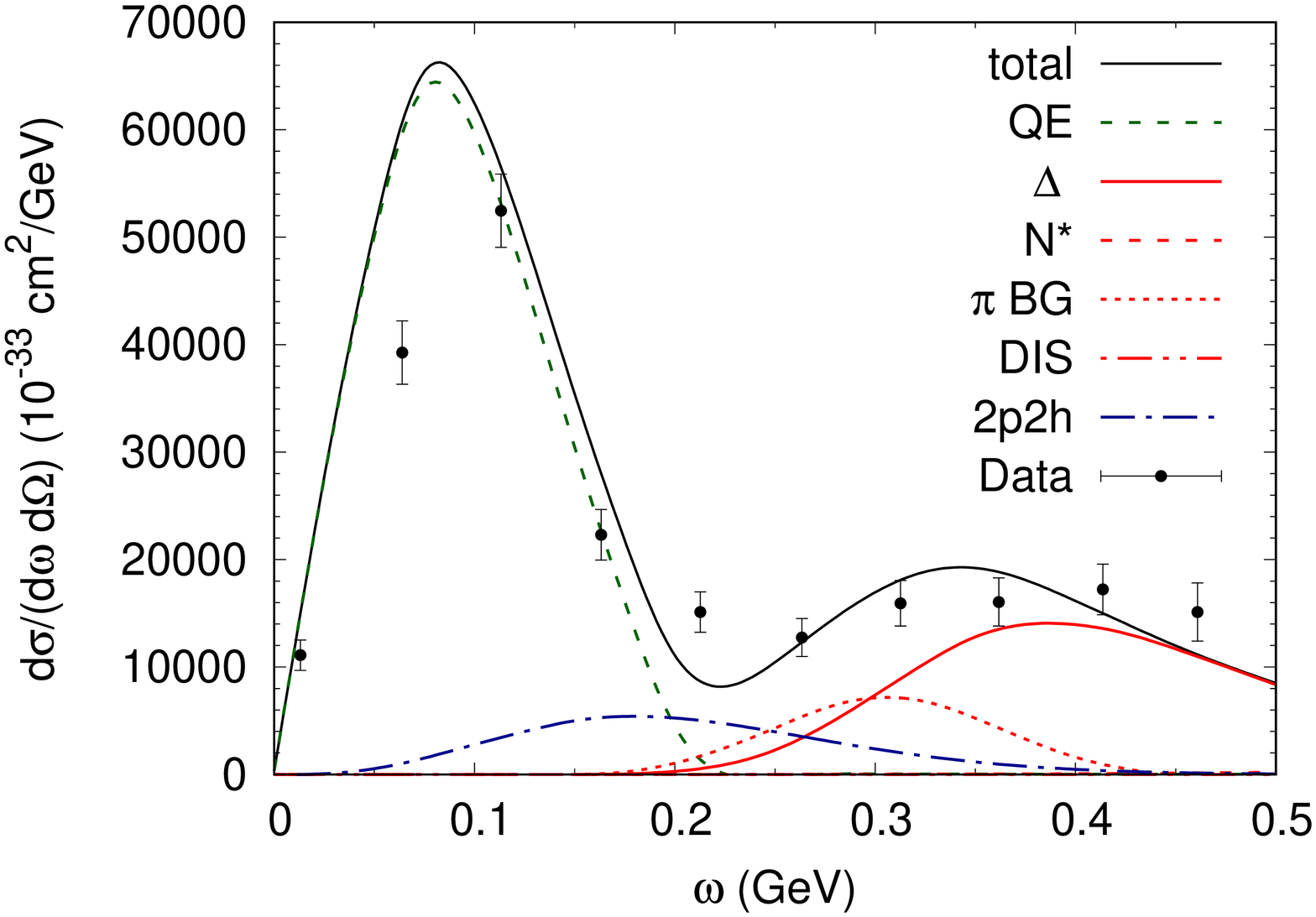}
\caption{E= 0.7 GeV, $\theta$= 32 degrees} \label{fig:a}
\end{subfigure}\hspace*{\fill}
\begin{subfigure}{0.5\textwidth}
	\centering
\includegraphics[width=0.5\linewidth,angle=-90]{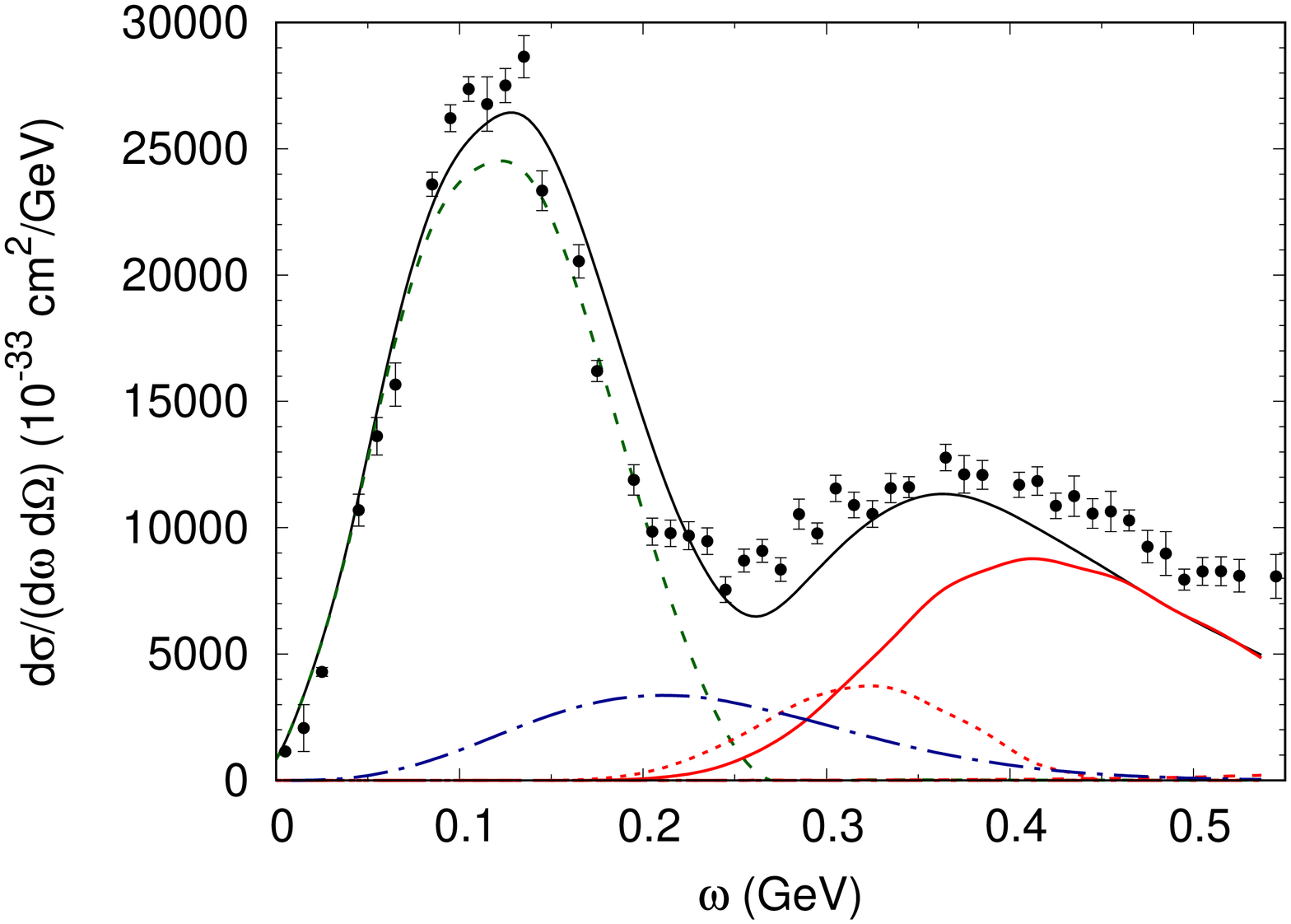}
\caption{E= 0.737 GeV, $\theta$ = 37.1 degrees} \label{fig:b}
\end{subfigure}
\medskip
\begin{subfigure}{0.5\textwidth}
	\centering
\includegraphics[width=0.5\linewidth,angle=-90]{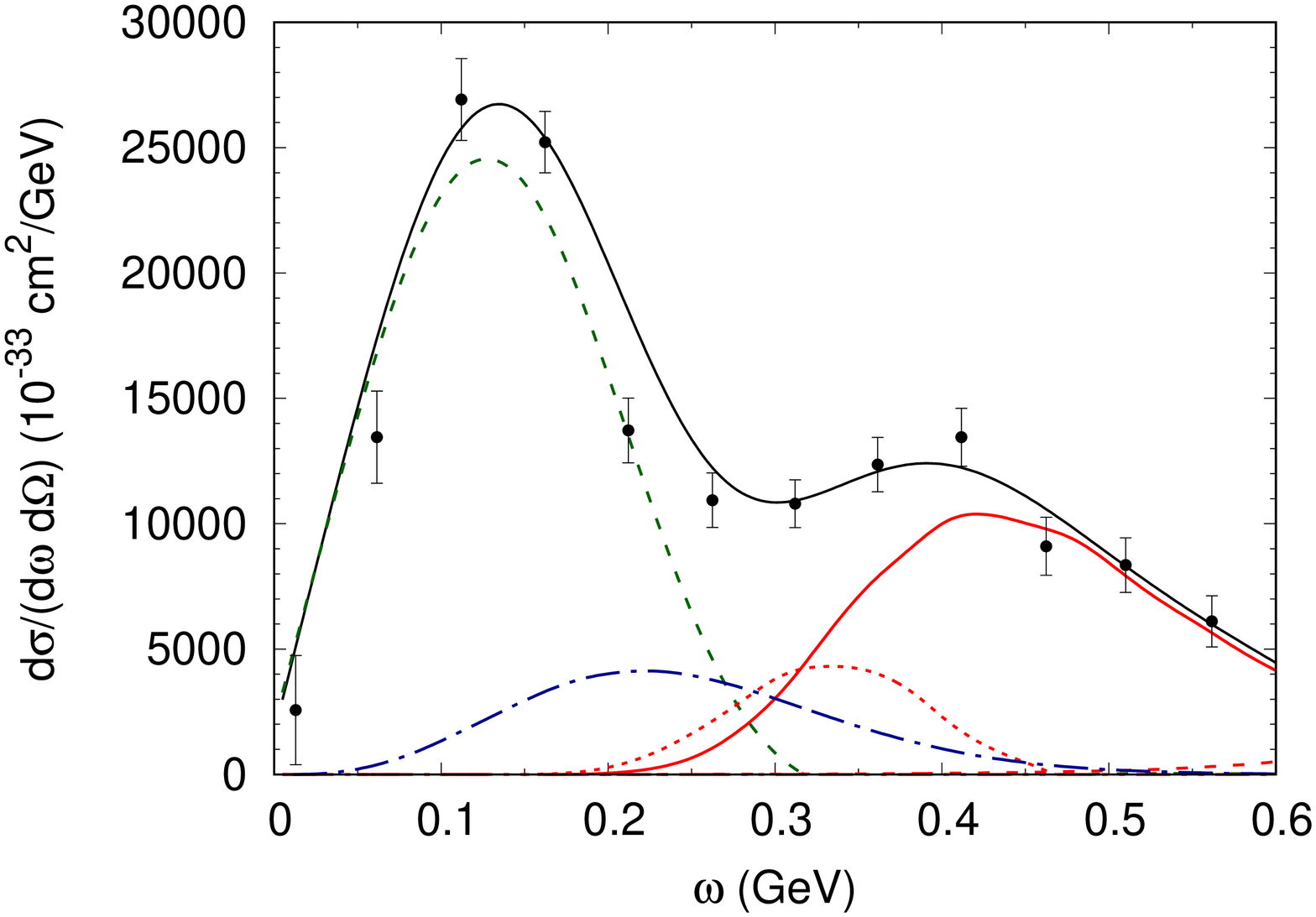}
\caption{E = 0.88 GeV, $\theta$ = 32 degrees} \label{fig:c}
\end{subfigure}\hspace*{\fill}
\begin{subfigure}{0.5\textwidth}
	\centering
\includegraphics[width=0.5\linewidth,angle=-90]{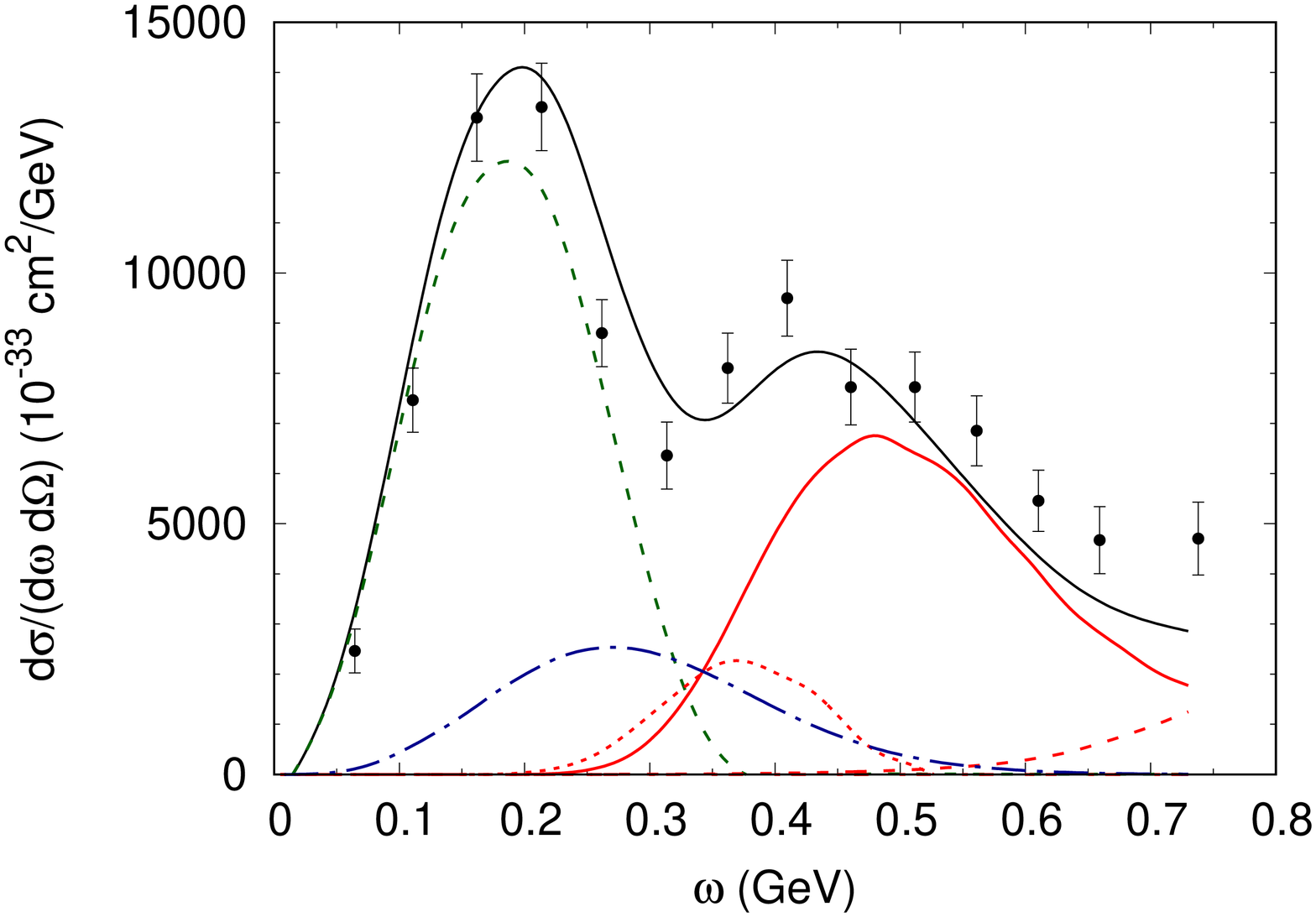}
\caption{E = 1.080 GeV, $\theta$ = 32 degrees} \label{fig:d}
\end{subfigure}
\medskip
\begin{subfigure}{0.5\textwidth}
	\centering
\includegraphics[width=0.5\linewidth,angle=-90]{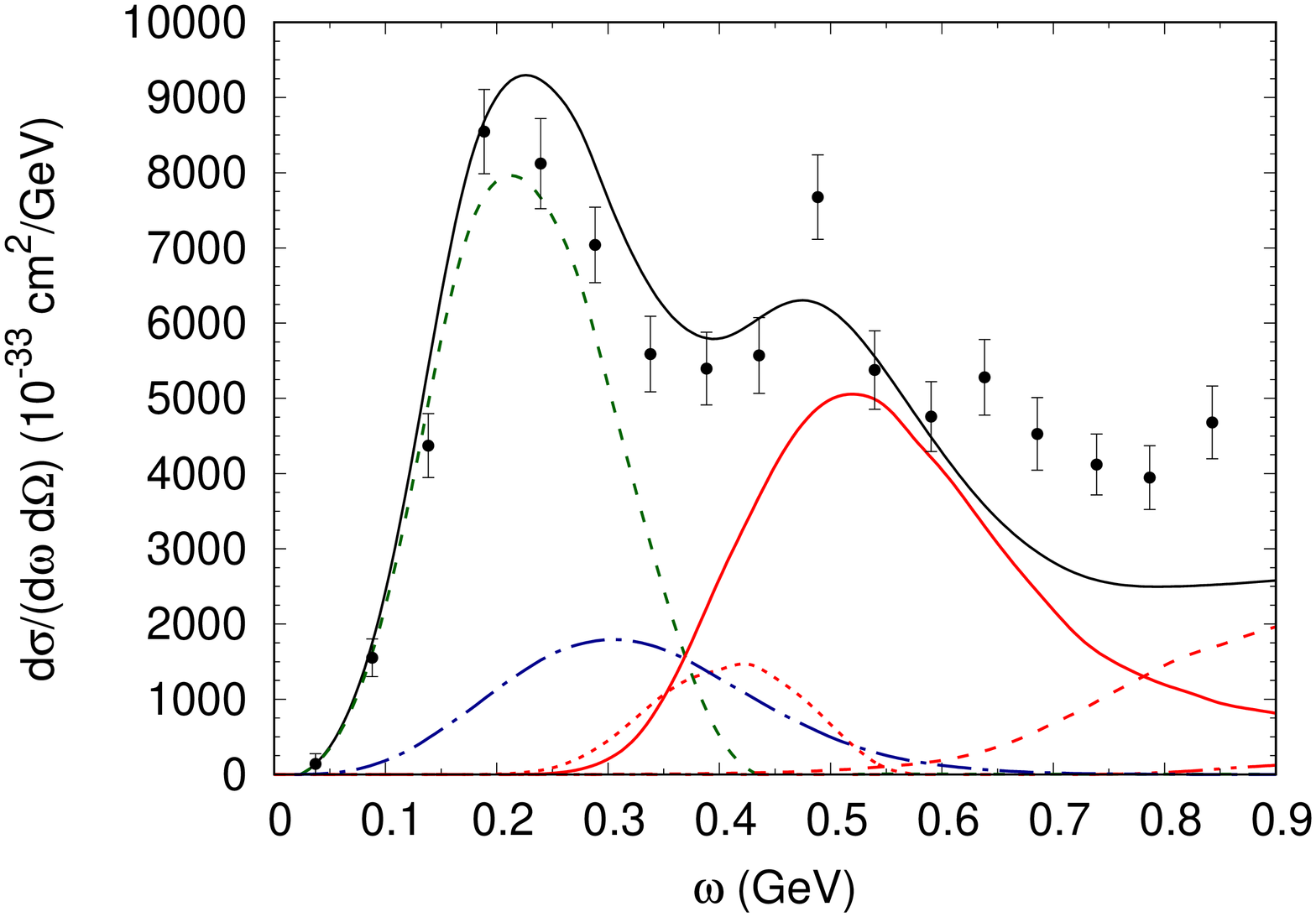}
\caption{E = 1.2 GeV, $\theta$ = 32 degrees} \label{fig:e}
\end{subfigure}\hspace*{\fill}
\begin{subfigure}{0.5\textwidth}
	\centering
\includegraphics[width=0.725\linewidth]{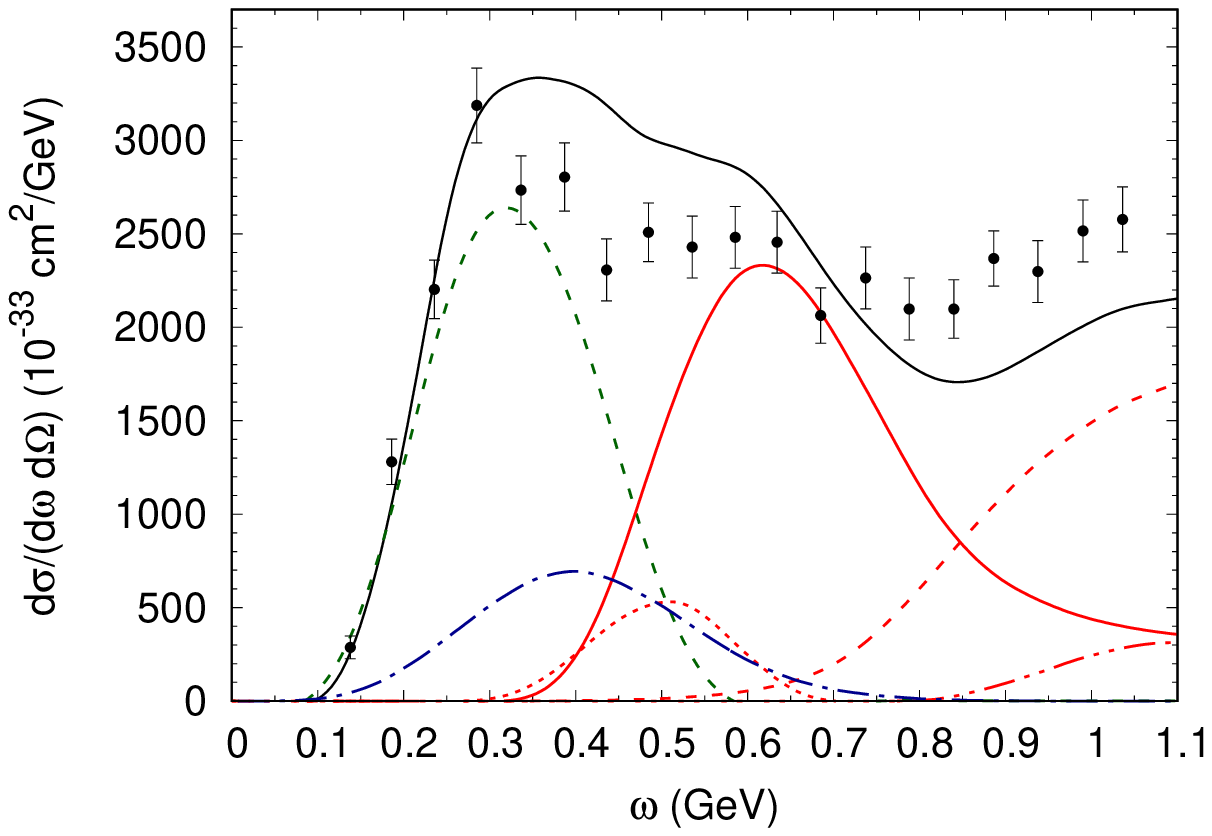}
\caption{E = 1.5 GeV, $\theta$ = 32 degrees} \label{fig:f}
\end{subfigure}
\caption{Inclusive cross sections for interactions of electrons with $^{16}O$. The electron energies and scattering angles are given below each figure.  The individual curves give the contributions of the subprocesses QE scattering, 2p2h interactions, $\Delta$ resonance contributions, contributions of higher-lying resonances $N^*$, the $1\pi$ background con\-tri\-butions and finally those from deep inelastic scattering. The data are taken from the arxive for quasielastic electron-nucleus scattering data \cite{Benhar:2006er}.}
\label{fig:eA}
\end{figure*}
In Fig.\ \ref{fig:eA} we show the results for all the measured inclusive cross sections for interactions with the target nucleus $^{16}$O, covering $Q^2$ values from $Q^2= 0.13$ GeV$^2$ (uppermost left) to $Q^2 = 0.55$ GeV$^2$ (lowest right). Both of the energy and the $Q^2$ ranges are relevant for the T2K experiment where the neutrino flux peaks at about 0.6 GeV.

In all the cases shown here the QE peak and the $\Delta$ peak overlap significantly at the high-energy side of the QE peak. In addition, there also the non-resonant pion background terms contribute. As in $^{12}$C the 2p2h contribution, while being peaked in the dip region between QE and $\Delta$ peaks, is present already at lower energies at the QE peak.

Overall, the calculations -- without any free parameter -- reproduce the various cross sections quite well; the highest energy (1.5 GeV) data show the largest discrepancy (up to $\approx 20$\%) in the QE peak region. It must be noted, however, that the absolute size of the cross sections falls quite significantly with energy. At the highest energy of 1.5 GeV it amounts to only about 1/20 of the values at the lowest energy of 0.7 GeV. Furthermore, the neutrino experiments average both over the incoming energy and over the energy transfer $\omega$ so that the weaker cross sections contribute less and particular structures in this inclusive electron cross section will get smeared out in $\nu A$ reactions.

A similar comparison for $^{16}$O was recently shown in Ref.\ \cite{Megias:2017cuh} based on the scaling model. For the four lowest energies the agreement with the data is comparable to the one obtained here while it is closer to the data at the higher energies. This agreement is reached there by using empirically determined scaling functions both in the QE and the resonance region. In addition a shift parameter and a Fermi-momentum are adjusted from nucleus to nucleus; the latter affects the 2p2h contribution. The 2p2h contributions are very similar in their magnitude to the ones obtained in the present calculations. They peak, however, at an energy transfer that is about 50 MeV higher than in the present calculations; in the work of Ref.\ \cite{Megias:2017cuh} this peak value is sensitive to the value of the Fermi-momentum fitted to this contribution.

\section{Neutrino-Nucleus interactions}
In this section we compare the GiBUU results with the recent neutrino data on events with zero pions on C and O targets in the T2K near detector \cite{Abe:2017rfw}.

\subsection{C target}
\begin{figure*}[t]
\centering
\includegraphics[width=0.41\textwidth,height=4.6cm]{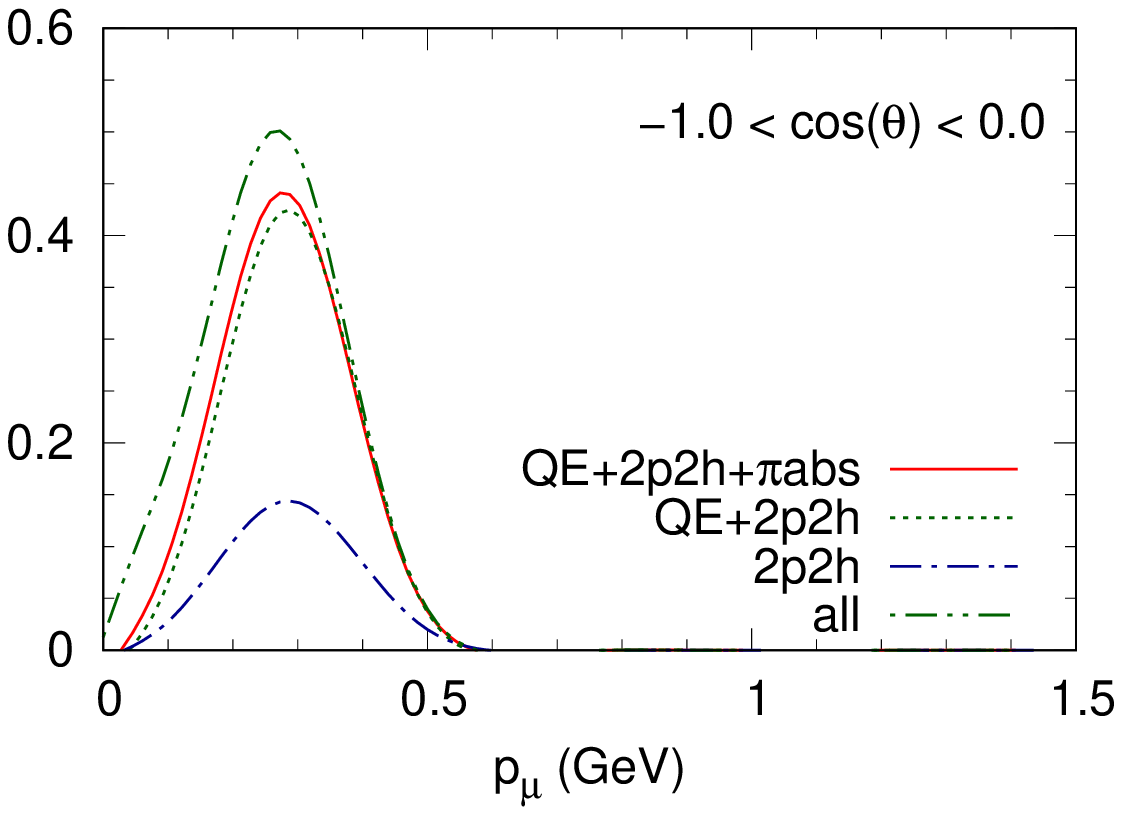}
\includegraphics[width=0.9\textwidth]{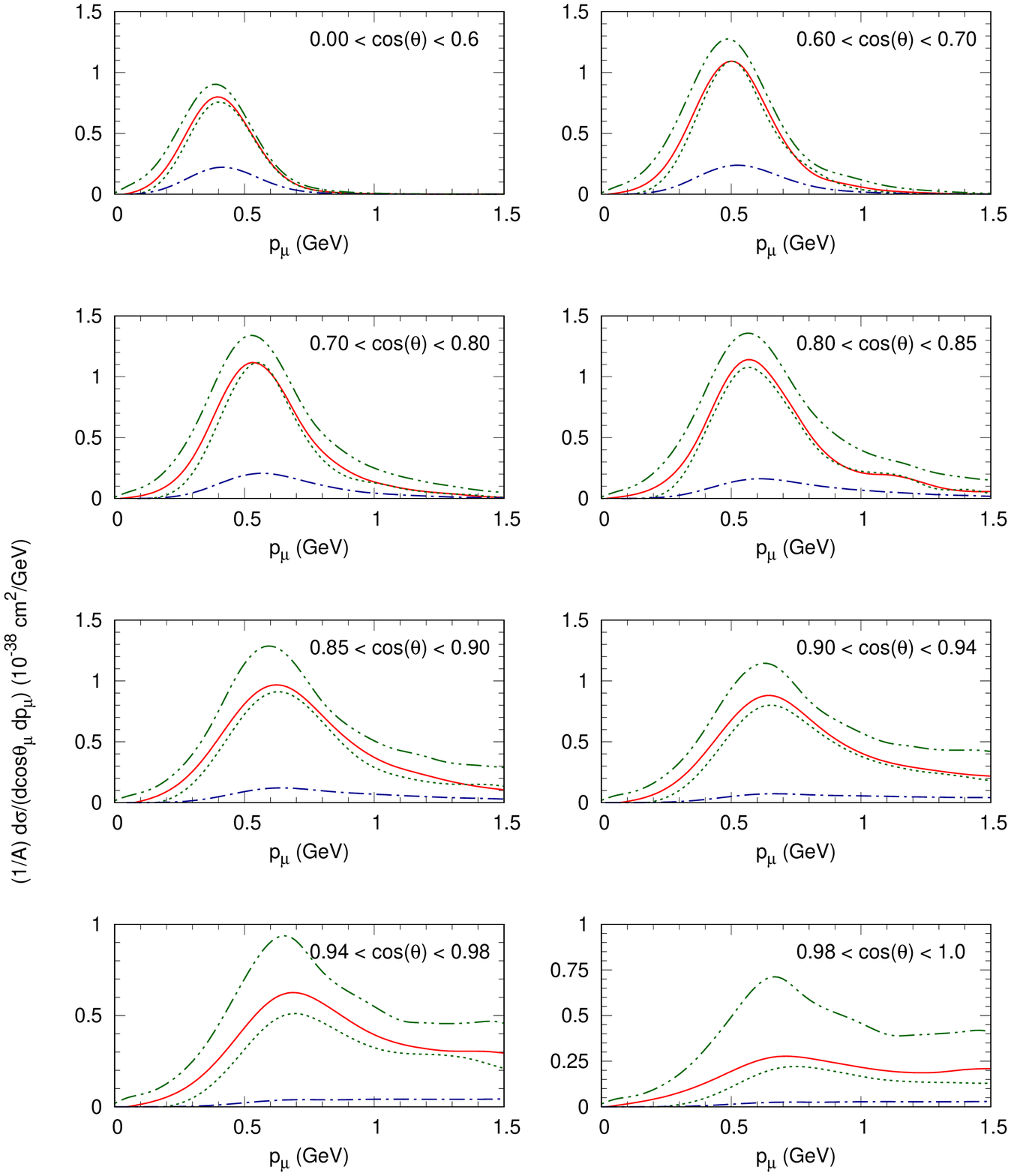}
\caption{Muon momentum distributions per nucleon in the various angular bins as used by the T2K analysis \cite{Abe:2016tmq} for a C target. The top dash-dot-dotted curve gives the fully inclusive cross section. The solid (red) curves give the total cross section for events with 0 pions in the outgoing channel. The dotted (green) curves give the cross sections for QE + 2p2h alone; the dashed-dotted (blue) gives the 2p2h contribution alone.}
\label{fig:Cpspectr}
\end{figure*}

In Fig.\ \ref{fig:Cpspectr} we first analyze the different contributions to the total cross section as a function of outgoing lepton momentum $p_\mu$. The subfigures in Fig.\ \ref{fig:Cpspectr}, corresponding to the experimental angular bins, show these momentum distributions for the primary reaction channels QE, 2p2h and ''stuck-pion'' events. The latter denote events in which pions were first produced and then subsequently reabsorbed. The highest curve gives the fully inclusive cross section in each angular bin. It is immediately seen that the explicit pion production, given by the difference between the highest and the second highest curves, increases when going forward. At the most backward angles it amounts to only about 15\% of the total whereas it becomes dominant at the most forward angles ($\approx$ 70\%).

\paragraph{2p2h absorption}
The contributions from 2p2h absorption processes are shown by the lowest, dash-dotted line in Fig.\ \ref{fig:Cpspectr}.  They are most essential at the backward angles due to the transverse character of this process. While at the backwards angles their contribution at the peak amounts to about 25\% of the total 0-pion cross section they become very small and very flat at forward angles.  At the most backwards angle the 2p2h contribution is peaked at $p_\mu \approx 0.4$ GeV; the peak moves to higher $p_\mu$ and becomes broader and less distinct with increasing $cos(\theta)$. This behavior is similar to the behavior found for the MiniBooNE double differential cross sections where the peak is at about 0.3 GeV at the back\-ward angles and then moves up to about 0.5 GeV at the most forward angular bin (see Fig.\ 5 in \cite{Gallmeister:2016dnq}).

\paragraph{Pion production and absorption}
While the solid line gives the cross section for all 0-pion events the dotted curves shows the cross section for QE + 2p2h events only. It is seen that a noticeable difference shows up only for $cos(\theta) > 0.7$, with the difference becoming the larger the more forward the scattering takes place. At the two most forward bins the difference amounts up to 20\% of the total 0-pion cross section at the peak. This reflects the fact that the pion production cross section on nuclei shows a distinct forward peaking as can directly be read off from comparing the topmost dash-dotted-dotted curve representing the fully inclusive cross section with the solid curve representing the cross section for 0-pion events (see also Fig.\ 5 in \cite{Mosel:2017nzk}. At the most forward angles the stuck-pion contribution is significantly larger than the 2p2h contribution.

\paragraph{Comparison with experiment}

\begin{figure*}[t]
\centering
\includegraphics[height=4.6cm, width=0.40\textwidth]{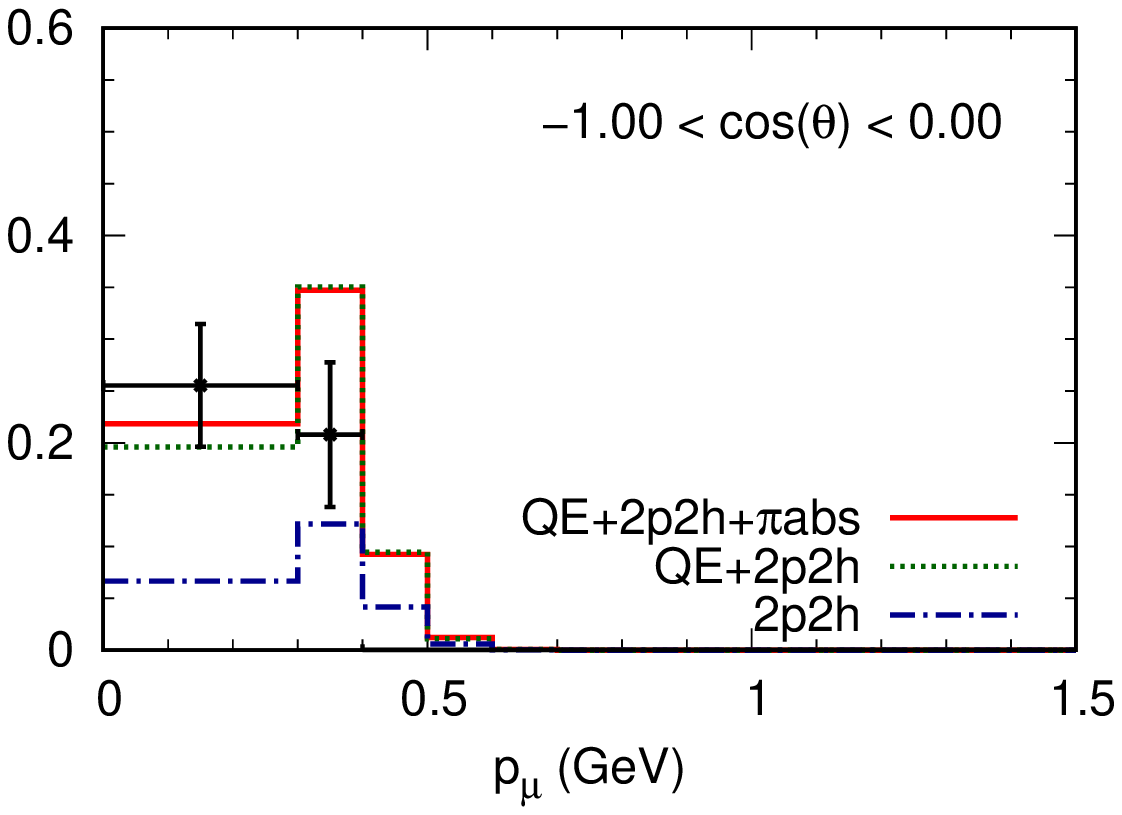}
\includegraphics[width=0.9\textwidth]{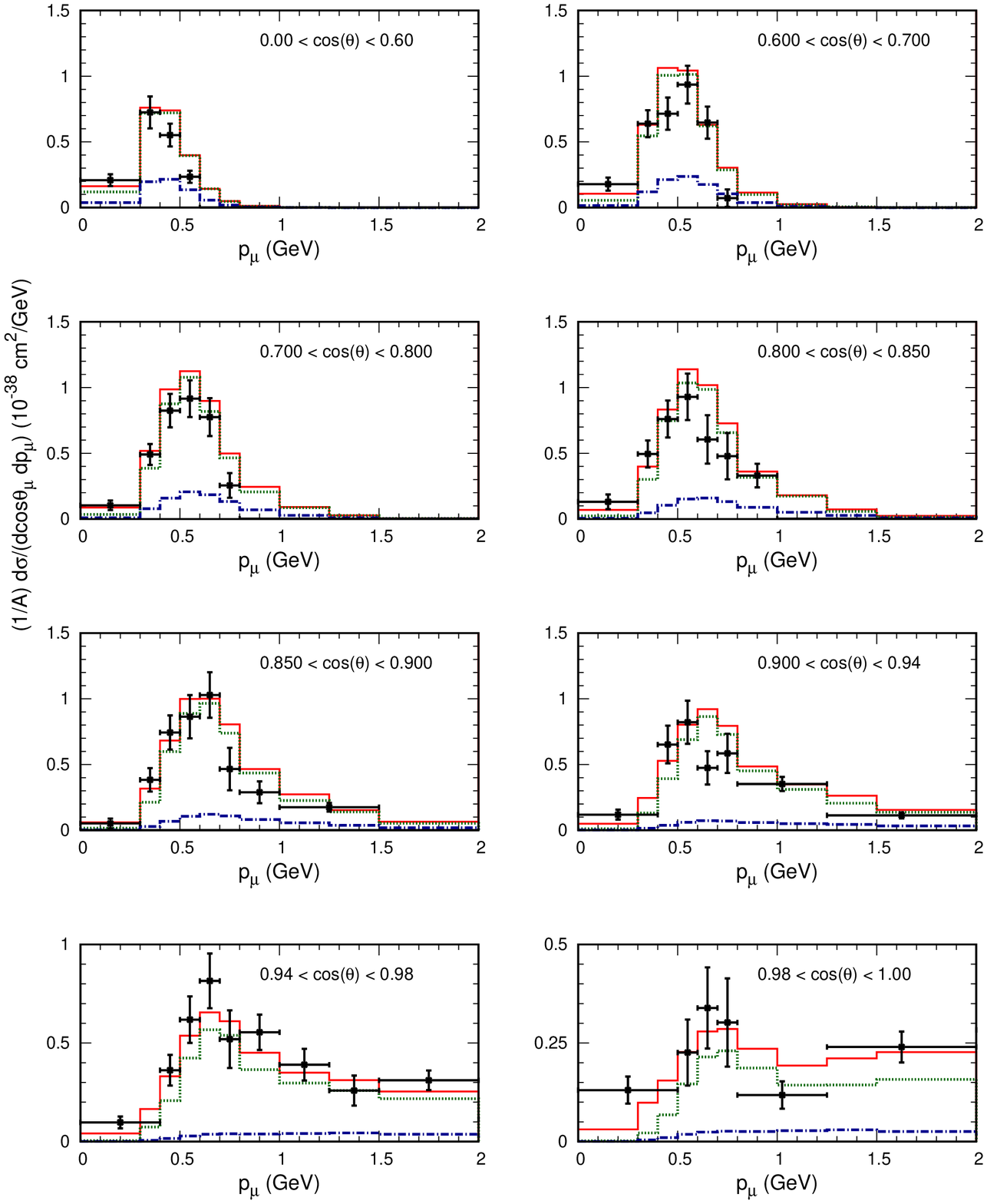}
\caption{Muon momentum distributions per nucleon  for 0-pion events in the various angular bins used by the T2K analysis \cite{Abe:2016tmq} for a C$_8$H$_8$ target. The solid (red) curves give the total cross section per nucleon for events with 0 pions in the outgoing channel for the isospin $\mathcal{T}=1$ 2p2h contribution. The dotted (green) curves give the cross sections for QE + 2p2h alone; the dashed-dotted (blue) gives the 2p2h contribution alone. The data are taken from \cite{Abe:2016tmq}.}
\label{fig:Cexphist}
\end{figure*}

Fig.\ \ref{fig:Cexphist} shows the momentum distributions of outgoing muons averaged over the experimental angular bins \cite{Abe:2016tmq} in comparison with experiment.
While the overall agreement with experiment is quite good there are some noticeable deviations which could possibly be used to fix theoretical uncertainties.

Closer inspection of Fig.\ \ref{fig:Cexphist} shows that the cross sections  in the most backward angular bin ($-1 < \cos(\theta) < 0$) are overestimated around a muon momentum of about 0.3 GeV. This is just where the 2p2h contribution has its maximum. Halving this contribution would bring the theory curve down to the upper boundary of the experimental error bar. A similar situation holds also for all the other bins up to $\cos(\theta) < 0.90$ where halving the 2p2h contribution would always lead to a better agreement with experiment. We note that also the calculations within the Martini and the Nieves models shown in Ref.\ \cite{Abe:2016tmq} exhibit a very similar disagreement in these bins where the stuck-pion events are not essential.

In \cite{Gallmeister:2016dnq} we have discussed that the transition from electron- to neutrino-induced structure functions involves a factor $\mathcal{T}+1$ where $\mathcal{T}$ is the target's isospin. The nearly perfect agreement with the MiniBooNE double-differential data was obtained there without any flux renormalization  with $\mathcal{T}$=1\footnote{The value $\mathcal{T} = 1$ was also used in all calculations reported in this paper.}. However, we have also discussed in \cite{Gallmeister:2016dnq} that a determination of $\mathcal{T}$=1 or =0  from the data, i.e.\, of a factor 2 in the strength of the 2p2h contributions, requires a flux determination to better than 10\% uncertainty. Halving the 2p2h contribution there would require an overall decrease of the incoming flux in the MiniBooNE by less than 10\% which is well within the experimental uncertainty.

\subsection{H$_2$O target}
\begin{figure*}[ht]
\includegraphics[width=0.9\textwidth]{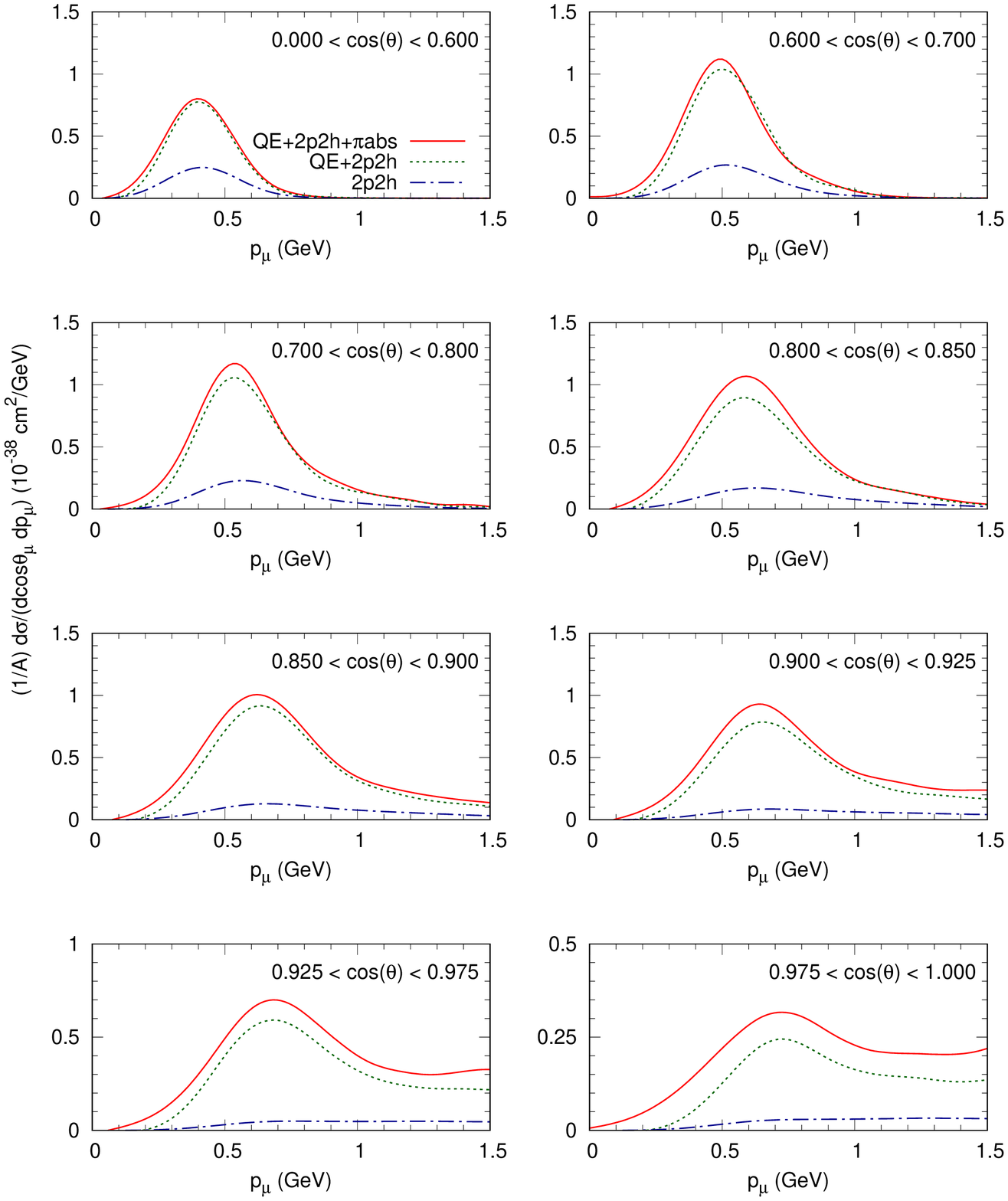}
\caption{Muon momentum distributions per nucleon in the various angular bins as used by the T2K analysis \cite{Abe:2017rfw} for an H$_2$O target. The solid (red) curves give the total cross section per nucleon for events with 0 pions in the outgoing channel. The dotted (green) curves give the cross section without the stuck-pion events and the dashed-dotted (blue) curves give the contribution of the 2p2h excitations.}
\label{fig:Opspectr}
\end{figure*}

The momentum spectra for this target are given in Fig,\ \ref{fig:Opspectr}. Comparing the results shown here with those for $^{12}$C in Fig.\ \ref{fig:Cpspectr} shows that the cross sections per nucleon are essentially identical for all components.

\paragraph{2p2h absorption}Again, the 2p2h contribution is most dominant at the more backwards angles. This reflects the transverse nature of the 2p2h contribution. At forward angles the 2p2h contribution becomes negligible.

\paragraph{Pion production and absorption}
While the solid line gives the cross section for all 0-pion events the dotted curves shows the cross section for QE + 2p2h events only. A noticeable difference shows up only for $cos(\theta) > 0.7$, with the difference becoming larger as more forward scattering takes place. At the two most forward bins, the difference, i.e.\ , the cross section for stuck-pion events, amounts to 20\% of the total 0-pion cross section at the peak.

\begin{figure*}[t]
\includegraphics[width=0.9\linewidth]{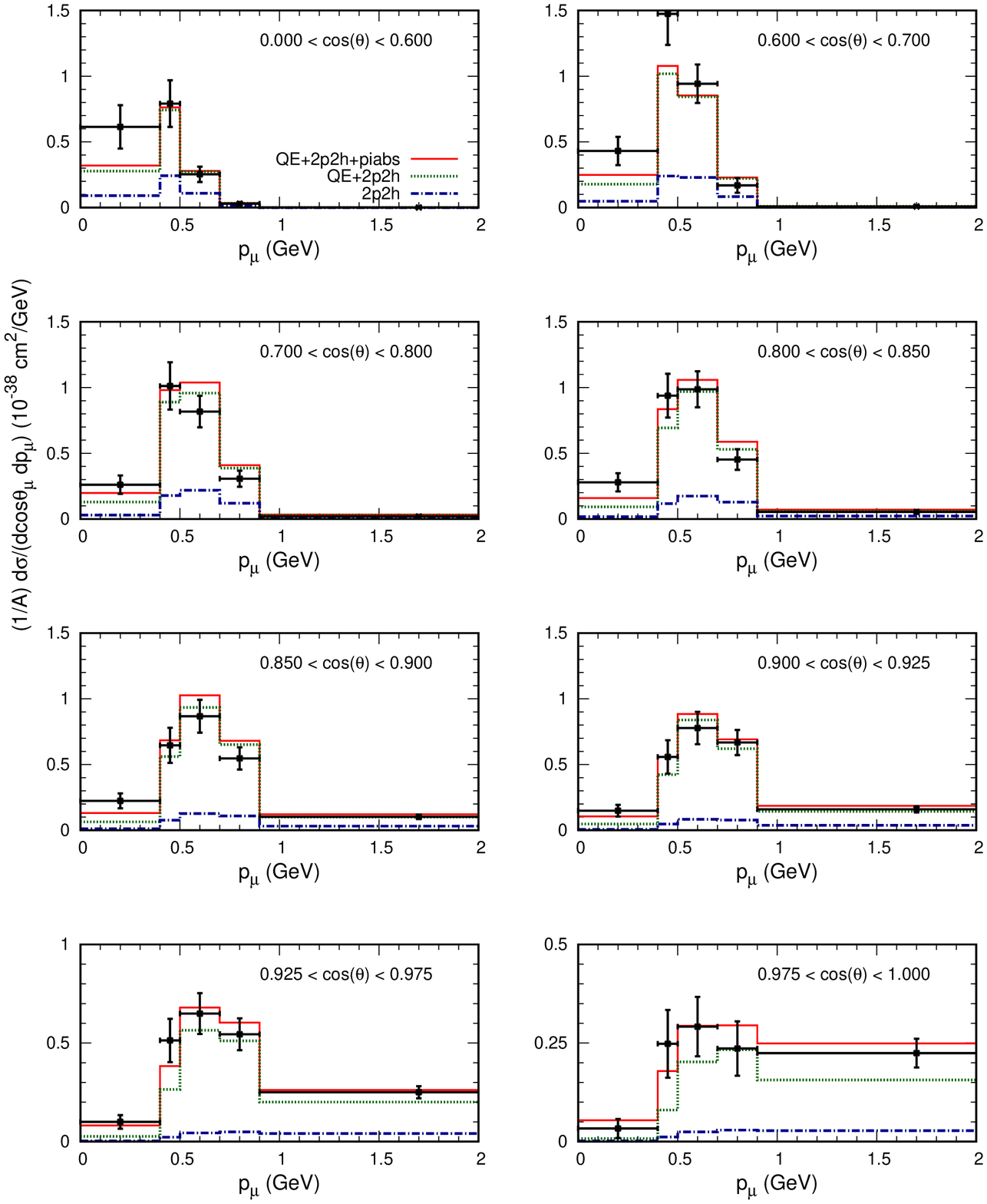}
\caption{Comparison of GiBUU results (step-like solid curve) with the H$_2$O 0-pion data from \cite{Abe:2017rfw}. The (green) dotted curve gives the cross section per nucleon for 0-pion events from QE and 2p2h processes only. The (blue) dashed-dotted line gives the contributions from 2p2h processes.}
\label{fig:binspectr}
\end{figure*}

\paragraph{Comparison with experiment}
In Fig.\ \ref{fig:binspectr} we now show the comparison of our calculations with experiment. For this comparison we have averaged the momentum distributions shown in Fig.\ \ref{fig:Opspectr} also over the corresponding momentum bins. It is seen that the calculations agree very well with experiment in all angle and momentum bins. The only significant disagreement shows up for the lowest momentum in the angular bin $0.0 < cos(\theta) < 0.6$ where the calculated value is lower than that obtained in experiment. This very same disagreement also shows up in all the comparisons with generators and theories given in Ref.\ \cite{Abe:2017rfw}. For the C target discussed earlier the experimental cross section is significantly lower in this bin than it is here for the O target; there, the theory reproduced the experimental value.

A closer inspection of the various angular bins shows that while the 2p2h contribution is largest at the largest angles and then decreases when going to forward angles, the stuck-pion events show an opposite behavior. At forward angles they are essential for good agreement with the experimental results. Contrary to the C case discussed earlier, here halving the 2p2h contribution would not lead to any significantly better agreement with experiment, but such modified calculation would still be in agreement with the data.

\subsubsection{Comparison with other theories}
The experimental paper \cite{Abe:2017rfw} already contains comparisons with calculations by the SUSA group and by Martini et al.\, the latter though only for $^{12}$C. The SUSA group has recently also published a more extensive paper on this reaction \cite{Megias:2017cuh}. We now compare our results with these model calculations.

\subparagraph{2p2h processes.} For the electron-induced reactions discussed earlier (see Fig.\ \ref{fig:eA}) the 2p2h contributions obtained by Megias et al.\ \cite{Megias:2017cuh} are very similar to the ones obtained here. It is, therefore, surprising to see that they differ significantly for the neutrino-induced reactions, both in magnitude and momentum-dependence. In Ref.\ \cite{Megias:2017cuh} the 2p2h contributions always contribute on the lower momentum side of the QE peak whereas in our calculations they are located in the peak region. Also, while for the electrons the overall magnitudes of the 2p2h contributions roughly agree with each other, they are quite different for neutrinos. For example, for the most forward bin $0.975 < cos(\theta) < 1.000$ in \cite{Megias:2017cuh} they amount to about 30\% of the total at large $p_\mu$ whereas they contribute significantly less ($< 10\%$) in our calculations. As a consequence, in Ref.\ \cite{Megias:2017cuh} the QE + 2p2h contributions alone already overestimate the experimental cross section in the peak region at forward angles by about 30\% (see Fig.\ 3 in Ref.\ \cite{Megias:2017cuh}).

In Ref.\ \cite{Abe:2017rfw} also results of a calculation within the Martini model for a $^{12}$C target are shown. These calculations also show a larger 2p2h contribution than our calculation, in particular in the most forward angular bins. As a consequence again the calculations for QE+2p2h alone overestimate the data there.

The authors of Ref.\ \cite{Megias:2017cuh} speculate that this disagreement at forward angles might be due to deficiencies in their treatment of Pauli-blocking. The cross section there is given by a sum of QE-scattering and 2p2h contributions. The former usually are quite insensitive to details as model comparisons have shown. We, therefore, speculate, that this disagreement could also be due to an overestimation of the longitudinal contribution to the 2p2h response. This interpretation is line with the GiBUU results, which on one hand describe the data also at forward angles quite well. On the other hand, these calculations do not contain a longitudinal 2p2h contribution thus leading to a significantly smaller overall 2p2h contribution at forward angles.

\subparagraph{Stuck pion events}
As shown earlier,in the present calculations the stuck-pion events contribute about 20\% of the total cross section in the most forward angular bin.
The 2p2h contribution is negligible there and the total agrees very well with experiment.

Stuck-pion events are partially contained in the calculations of Megias et al.\ \cite{Megias:2017cuh} and those of Martini et al.\ \cite{Abe:2017rfw} because both of these approaches involve internal off-shell pion lines\footnote{We note that such processes are also contained in the fsi encoded in GiBUU}. However, the reabsorption of asymptotic on-shell pions is missing in both of these approaches. Nevertheless both of them already overestimate the cross section at forward angles quite significantly. In both models the 2p2h contribution is responsible for this overestimate. Adding the on-shell stuck-pion contribution to their result would overestimate the forward cross section even further.

\section{Summary}
Pion production is an essential process in neutrino-nucleus interactions and thus has to be under quantitative control. This applies both to reactions with explicit pions in the final state and to reactions with 0 pions in the outgoing channel. GiBUU describes the explicit pion production on  CH target and H$_2$O targets in the T2K beam.  It is, therefore, gratifying to see that this theory (and code) also describe the 0-pion events on both targets. For this result it was essential to use a theory that contains all three ingredients that contribute to the measured 0-pion cross sections: QE-scattering, 2p2h interactions and pion production and absorption.

The comparison with the 0-pion data on C shows that reducing the 2p2h contribution by a factor of 2, corresponding to a smaller target isospin, would improve the agreement with experiment at some of the not-so-forward angles. On the other hand, the agreement with the 0-pion data on O is already quite good, but halving the 2p2h contribution would also not deterioate the agreement significantly. The present experimental uncertainties thus do not allow to pin down the 2p2h contribution within a factor of 2.

Our calculations show that at forward angles the stuck-pion events contribute significantly to the 0-pion cross section. The T2K collaboration chose to compare their data with two theoretical models that lack the full description of these stuck-pion events \cite{Abe:2017rfw}. The more detailed comparison of SUSA calculations with the data by Megias et al.\ \cite{Megias:2017cuh} also suffers from the same deficiency. In this paper we have shown that the neglect of stuck-pion events can be justified only for back\-ward angles. For very forward angles this component is even larger than the 2p2h contribution. Thus, tuning a generator that does not contain this reaction channel to data leads to incorrect information about the other reaction channels (in this case mainly 2p2h).

This case also illustrates the limitation of all theories that can only describe inclusive cross sections; ab initio nuclear structure calculations, spectral function methods and SUSA all suffer from this problem. Even if they are generalized to include inelastic excitations they can be compared to experiment only for fully inclusive cross sections, but not for selective event classes, such as 0-pion events.

\appendix*
\section{MiniTutorial}
A MiniTutorial that describes how the results shown in this paper for $^{16}$O were obtained with GiBUU can be found at \url{https://gibuu.hepforge.org/trac/wiki/jobCards} as Example 10. There also the jobcards are available for download. They may serve as an example for similar studies with different fluxes and different target materials.

\begin{acknowledgments}
The authors are grateful to Teppei Katori and to Tianlu Yuan for helpful discussions on the experimental analysis. They also thank Jan Sobczyk for discussions on the correct treatment of experimental error bars.
\end{acknowledgments}

\bibliographystyle{apsrev4-1}
\bibliography{nuclear}

\end{document}